\documentclass[manuscript]{aastex}

\shorttitle{The Effect of Star Spots on Radius Determination of The Low Mass
 Binary GU Boo}
\shortauthors{Windmiller et al.}

\begin{document}

\title {The Effect of Star Spots on Accurate Radius Determination of \\
       the Low Mass Double-lined Eclipsing Binary GU Boo}

\author{G. Windmiller}
\affil{San Diego State University, San Diego, CA 92182-1221}
\email{windmill@rohan.sdsu.edu}

\author{J. A. Orosz}
\affil{San Diego State University, San Diego, CA 92182-1221}
\email{orosz@sciences.sdsu.edu}

\and

\author{P. B. Etzel}
\affil{San Diego State University, San Diego, CA 92182-1221}
\email{etzel@sciences.sdsu.edu}


\begin{abstract}
GU Boo is one of only a relatively
small number of well studied double-lined eclipsing
binaries that contain low-mass stars.  L\'opez-Morales
\& Ribas (2005, hereafter LR05)
present a comprehensive analysis of multi-color light
and radial velocity curves for this system.
The GU Boo light curves presented in
LR05 had substantial asymmetries, which were
attributed to large spots. In spite of the asymmetry LR05
derived masses and radii accurate to $\simeq$ $2\%$.
We obtained additional
photometry of GU Boo using both a CCD and a single-channel photometer
and
modeled the light curves with the $\rm ELC$ software to determine
if the large spots in the light curves give rise to systematic errors at
the few percent level. We also modeled the original
light curves from LR05 using models with and without spots.
We derived a radius of the primary of
$0.6329\pm 0.0026\,R_{\odot}$,
$0.6413\pm 0.0049\,R_{\odot}$, and
$0.6373\pm 0.0029\,R_{\odot}$ from the CCD, photoelectric, and
LR05 data, respectively.  Each of these measurements agrees  with
the value reported by LR05  ($R_1=0.623\pm0.016\,R_{\odot}$)
at the level of $\approx 2\%$.  In addition,
the spread in these values is $\approx 1-2\%$
from the mean.
For the secondary, we derive radii of
$0.6074\pm 0.0035\,R_{\odot}$,
$0.5944\pm 0.0069\,R_{\odot}$, and
$0.5976\pm 0.0059\,R_{\odot}$ from the three respective data sets.
The LR05 value is $R_2=0.620\pm 0.020\,R_{\odot}$, which is
$\approx 2-3\%$ larger than each of the three values
we found.  The spread in  these values is $\approx 2\%$ from the mean.
The systematic difference
between our three determinations of the secondary radius and that of
LR05 might  be attributed to differences in the modelling process
and codes used.
Our own fits  suggest that, for GU Boo at least,
using accurate spot modelling of a single set of multicolor
light curves
results in radii determinations accurate
at the $\approx 2\%$ level.
\end{abstract}


\section{Introduction}

Understanding the structure and evolution of stars is a
basic goal of stellar astrophysics, and is also
required in most other branches of astrophysics.
Detailed models of stellar evolution can predict (among other things)
the stellar radius as a function of mass and age.
The Sun can be used to calibrate stellar evolution models
since its mass, radius, and age are well determined (e.g., Guenther et
al.\ 1992).  Critical tests of evolution theory for stars other than
the Sun can be made on a small set of eclipsing binary stars (e.g.,
Pols et al.\ 1997; Schr\"oder et al.\ 1997).
For this purpose, it is essential to derive accurate masses and radii
for these binaries.

In general, the results of stellar evolution models
compare favorably to data for main sequence
stars with masses $\gtrsim 1\,M_{\odot}$
(e.g.\ Pols et al.\ 1997).  However,
the models for stars on the lower main sequence
have problems matching precise data from eclipsing binaries.
A good example is the double-lined eclipsing M-star binary YY Gem,
which is a member of the Castor group.  This binary contains
a pair of nearly identical stars with masses of $M=0.599\,M_{\odot}$.
Torres \& Ribas (2002) have shown that all models for stars on the
lower main sequence underestimate the radii of the YY Gem
components by up to 20\% and that most models overestimate the
effective temperatures by 150 K or more.  Similar trends are found in V818 Tau
(Torres \& Ribas 2002), CU Cnc (Ribas et al. 2003), GU Boo
(L\'opez-Morales \& Ribas 2005), TrES-Her0-07621 (Creevey et al.\, 2005),
 2MASS J05162881+2607387  Schuh et al. 2003;  Bayless \& Orosz, 2006),
NSVS 02502726 (\c{C}akirli et al.\ 2009), and
GJ 3236 (Irwin et al.\ 2009).
The disagreement between the models and the data for these binaries
is very troubling since models for low-mass stars are often used
to estimate the ages for open clusters and for individual T Tauri stars
by placing them in an H-R diagram.
There have been recent suggestions that
unusually strong stellar activity in
these low-mass stars
might make them larger than they otherwise would be
(Ribas 2006; Torres et al.\ 2006;
L\'opez-Morales 2007; Chabrier et al.\ 2007), and that the changes in the
stars caused by stellar activity has not been properly accounted
for in the evolutionary models.

In this paper, we focus on GU Boo.
The GU Boo light curves presented by L\'opez-Morales
\& Ribas (2005, hereafter LR05) are very precise, which allowed them to
derive radii accurate to a few percent using the well-known
Wilson \& Devinney (1971, W-D) code.
However, their light curves
are not symmetric about the primary eclipse.  The system is brighter
just before the secondary eclipse (i.e., at first contact)
than it is just after the secondary eclipse (i.e.,  at fourth contact).
The source of the excess light before secondary eclipse was attributed
by LR05 to a bright spot and a larger dark
spot on the primary. Although the W-D code can include spots, the spot model
is somewhat simplistic (e.g., one or two circular regions with a different
temperature than the rest of the star) and one has to wonder if
the spots cause systematic errors in the fitting at the few percent level.
One simple test of the robustness of the GU Boo parameters
given by LR05 is to obtain new light curves
and model all data independently.  In what follows, we present new
CCD and photoelectric observation of GU Boo obtained using the 1m
telescope at Mount Laguna Observatory.  In addition,
the radial velocities published by LR05 and the light curves
were kindly sent to us by Mercedes L\'opez-Morales.
We model our new light curves and the LR05 light curves
using various assumptions about the spots and compare our
results with those of LR95.
We end with a brief discussion and summary.

\section{OBSERVATIONS}

We  observed GU Boo during May-June of 2005 using
the 1m telescope at the Mount Laguna Observatory.  GU Boo was observed
with a Fairchild CCD 447,
backside illuminated $2048\times 2048$ with 15 micron
square pixels, and $V$, $R$, and $I$ filters. Other data were taken with
a single-channel photometer, employing an RCA C31034A GaAs based
photomultiplier, and $B$, $V$, $R$, and $I$ filters
(Bessell 1990). Table \ref{tab1} gives a summary of the observations.
Standard CCD image reductions were done in
IRAF\footnote{IRAF is distributed by the National Optical Astronomy
 Observatory, which are operated by the Association of Universities for
 Research in Astronomy, Inc., under the cooperative agreement with the
 National Science Foundation}.
The differential
light curves of GU Boo were derived using simple aperture photometry in
IRAF, including Stetson's curve-of-growth technique (Stetson 1990) to
derive optimal instrumental magnitudes corresponding to the largest aperture.
The reductions for the photoelectric data (hereafter PMT)
were done with the code FOTOM, which was
developed at San Diego State University. The new CCD light curves are
shown in Figure \ref{CCDlc}.  The PMT light curves are shown
in Figure \ref{PMTlc}.

\section{METHOD AND RESULTS}
\subsection{Light Curve Comparison}

\begin{figure}
\epsscale{.75}
\plotone{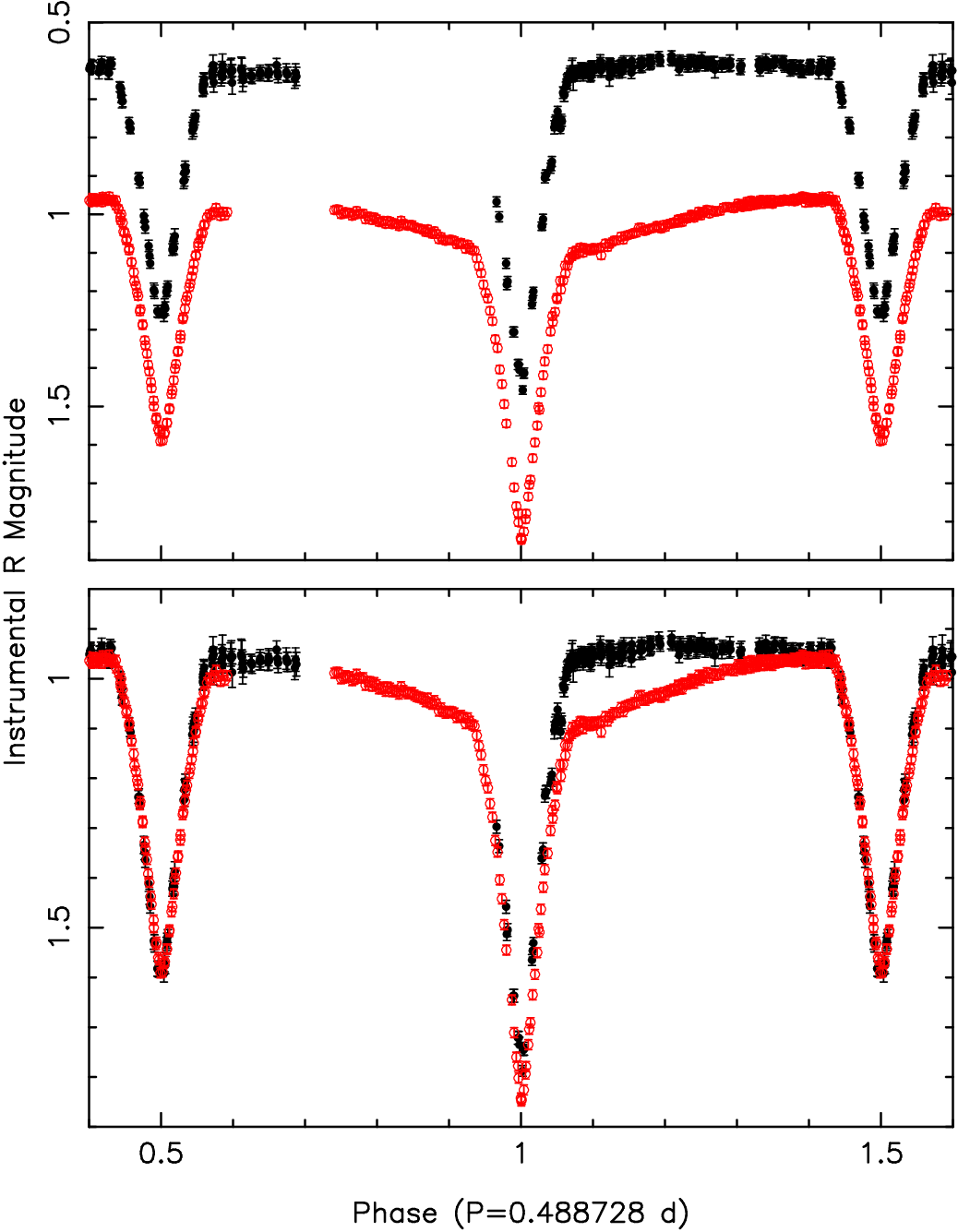}
\caption{Bottom:
A comparison of $R$-band data of GU Boo.  The
light curve from LR05 is shown with the
red open circles.
The CCD light curve obtained from Mount Laguna is shown with the
black filled
circles.
Top: Same as the bottom, but with the light curves
offset for clarity.
\label{complc}}
\end{figure}

Figure \ref{complc} compares our $R$-band CCD light curve
with the $R$-band light curve from LR05.
The light curves are rather different. As noted above, the light
curve from LR05 has a relatively large amount
of variation in the out-of-eclipse phases, exhibiting two different
slopes on either side of the primary eclipse.
In contrast, the light curves from Mount Laguna
are relatively
flat between eclipses and are more symmetric.
The secondary eclipse profiles are very similar, whereas the primary eclipse,
as well as the nearby phases, are depressed to fainter levels in the
LR05 light curve.
The natural interpretation is that there was
a rather large and dark spot on the primary when LR05
obtained their data, and that spot had mostly vanished by the time
we observed the system from Mount Laguna.  Indeed, LR05
invoked a large dark spot on the primary in their light curve
modeling. The source changed relatively little in the few weeks between
the CCD observations and the photoelectric observations (see
Figures \ref{CCDlc} and \ref{PMTlc}), making the spot(s)
stable on a time-scale
of a few weeks, which simplifies the light curve modeling discussed below.
As shown here, the existence and asymmetry of spots on either the
primary, the secondary, or both, can explain the differences between the two
light curves in both the asymmetry and the non-zero derivative
outside of eclipses.

\subsection{Light Curve Modeling}

We modeled our GU Boo light curves using our ELC code
(Orosz \& Hauschildt 2000) with updated model atmospheres for
low mass stars and brown dwarfs (Hauschildt, private communication). As noted
above,
the radial velocities published by LR05 and the light curves
were kindly sent to us by Mercedes L\'opez-Morales,
and we included
the radial velocities in all of our modeling runs.
The CCD light curves were modeled separately from
the PMT light curves.
Using ELC's various optimizers, we fit for the
following 13 parameters (the ranges searched are given in parentheses):
the primary mass $M_1$ ($0.5-0.7\,M_{\odot}$),
the primary radius $R_1$ ($0.57-0.67\,R_{\odot}$),
the ratio of the radii $R_1/R_2$ ($0.9-1.1$),
the inclination $i$ ($75-90^{\circ}$),
the $K$-velocity of the primary $K_1$ ($130.0-155.0$ km s$^{-1}$),
the effective temperatures $T_1$ ($3700-3900$~K) and
$T_2$ ($3600-3800$~K),
the orbital period $P$ ($0.48871-0.48874$ d),
the time of primary eclipse $T_0$ (HJD $2,452,723.98-2,452,723.99$),
and 4 parameters to describe a spot, namely the ``temperature
factor''\footnote{The temperature factor is the ratio between
 the temperature of the spot, $T_{\rm spot}$,
and the Effective temperature of the star, $T_{\rm eff}$, i.e.
$T_{f}={T_{\rm spot}/T_{\rm eff}}$} ($0.5-2.0$),
the latitude
($0-180^{\circ}$), the longitude
($0-360^{\circ}$), and the angular radius ($0-60^{\circ}$).
The mass ratio $Q$, and the semi-major axis $a$ are
mapped to directly from the fitted parameters $M_{1}$ and $K_{1}$.
We also modeled the LR05 light curves, for comparison, and
as an independent check on their results.

The radial velocity curves presented in LR05 have several observations
taken during secondary eclipse.  Curiously, the radial velocities of
the secondary star do not show any significant deviation from a sine
curve during the eclipse.  One normally observes a distortion in the
radial velocity curve during an eclipse (e.g., the Rossiter effect)
because of asymmetries in the absorption line profiles caused by the
partial covering of the star during a partial eclipse.  During some
initial model fits, it was found that the model radial velocity curve
for the secondary all had a large Rossiter effect.  Since the observed
velocity curve has a very small (if any) Rossiter effect, the fits to
the curve had larger $\chi^2$ values.  Since we are mainly using the
radial velocity curves to provide the scale of the binary and the mass
ratio, we excluded the radial velocities of the secondary that were
taken during the secondary eclipse.

For the CCD, PMT, and LR05 data sets, we modeled the data using ELC
for six different spot scenarios:
no spots, a single spot on the primary, a single spot on
the secondary, two spots on the primary, two spots on the secondary, and
one spot on each. Every one of these  cases involved the extensive use of
the ELC genetic optimizers, and the best-fitting model
was arrived at through iteration.  First, ELC's genetic optimizer
code was run
for a few hundred generations until convergence was reached.
As is often the case with modeling, the total
$\chi^2$ of the fit was larger than the number of data points.
The uncertainties on the measurements were scaled so that
the reduced $\chi^2$ was unity {\em for each bandpass and velocity
curve separately}. After the error bars were rescaled, the genetic
code was used again for several hundred more generations.  Next,
ELC's
Monte Carlo Markov Chain optimizer was  run several times, using
both random initial guesses and initial guesses supplied by the genetic code.
Finally, ELC's ``amoeba'' optimizer (an optimizer that uses a downhill
simplex method, see Press et al.\ 1992) was run, using as the initial
guess the best solution found from the genetic and Markov chain runs.
After the best solution was found,
we used the procedure outlined in Orosz et al.\ (2002)
to find approximate 1$\sigma$ confidence interval.
To estimate uncertainties on fitted and derived parameters, we projected the
multi-dimensional $\chi^2$ function into each parameter of interest.
The 1$\sigma$ confidence limits
was taken to be the ranges of the parameter where
$\chi^2 \leq \chi^2_{\rm min} + 2$.
Since the genetic ELC code samples parameter space
near $\chi^2_{\rm min}$ extensively, computing these limits is simple.
ELC saves from every computed model the $\chi^2$ of the fit, and the value
of the free parameters (e.g., the primary star mass, the ratio of the
radii, etc.), and the astrophysical parameters (e.g., the secondary star mass).
One can then choose the value of the parameter of interest at each value
of the $\chi^{2}$. We believe this method of uncertainty estimation is more robust
than the probable errors reported by W-D, although at the expense of
considerably more computer time.

\section{DISCUSSION}

\begin{figure}
\epsscale{.75}
\plotone{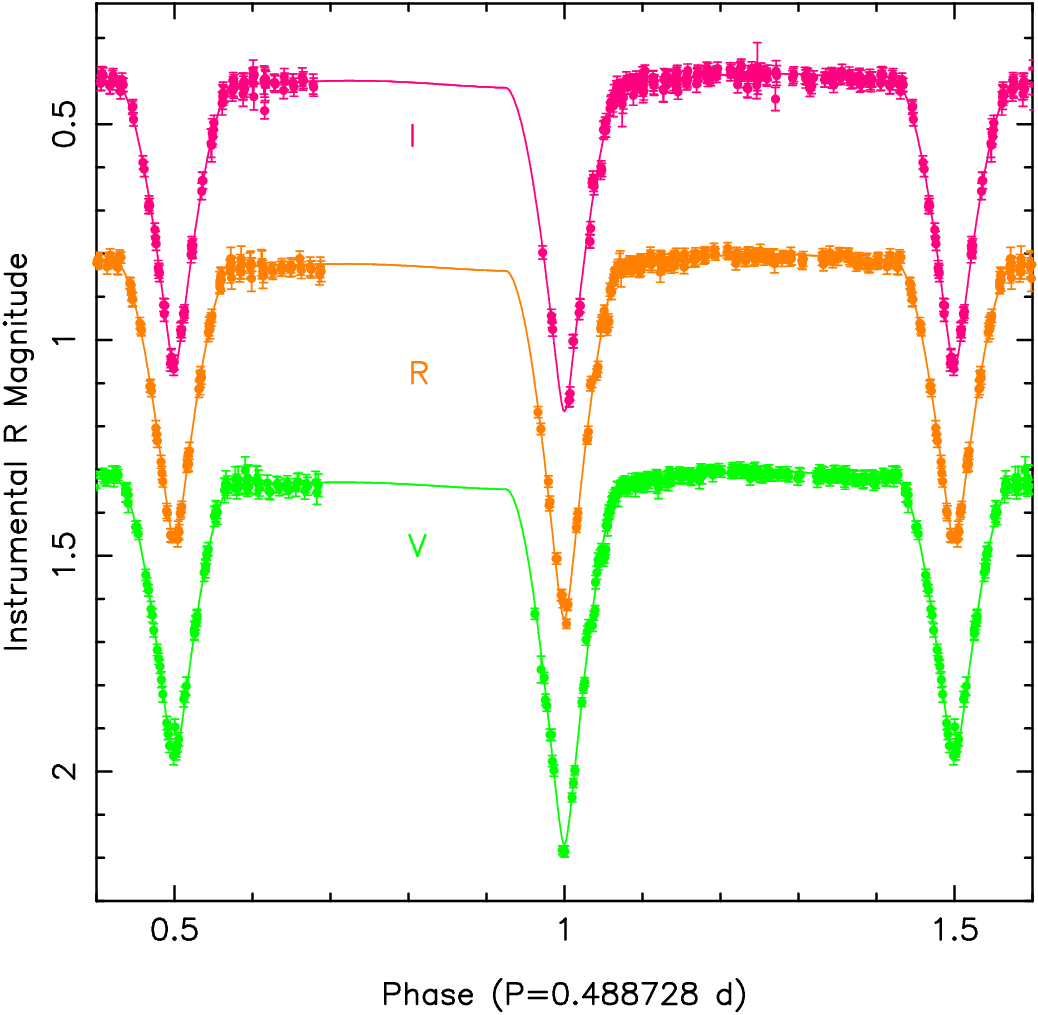}
\caption{The CCD light curves of GU Boo
obtained from Mount Laguna for the $V$,
$R$, and $I$ filters and the best-fitting ELC model.
\label{CCDlc}}
\end{figure}

\begin{figure}
\epsscale{0.75}
\plotone{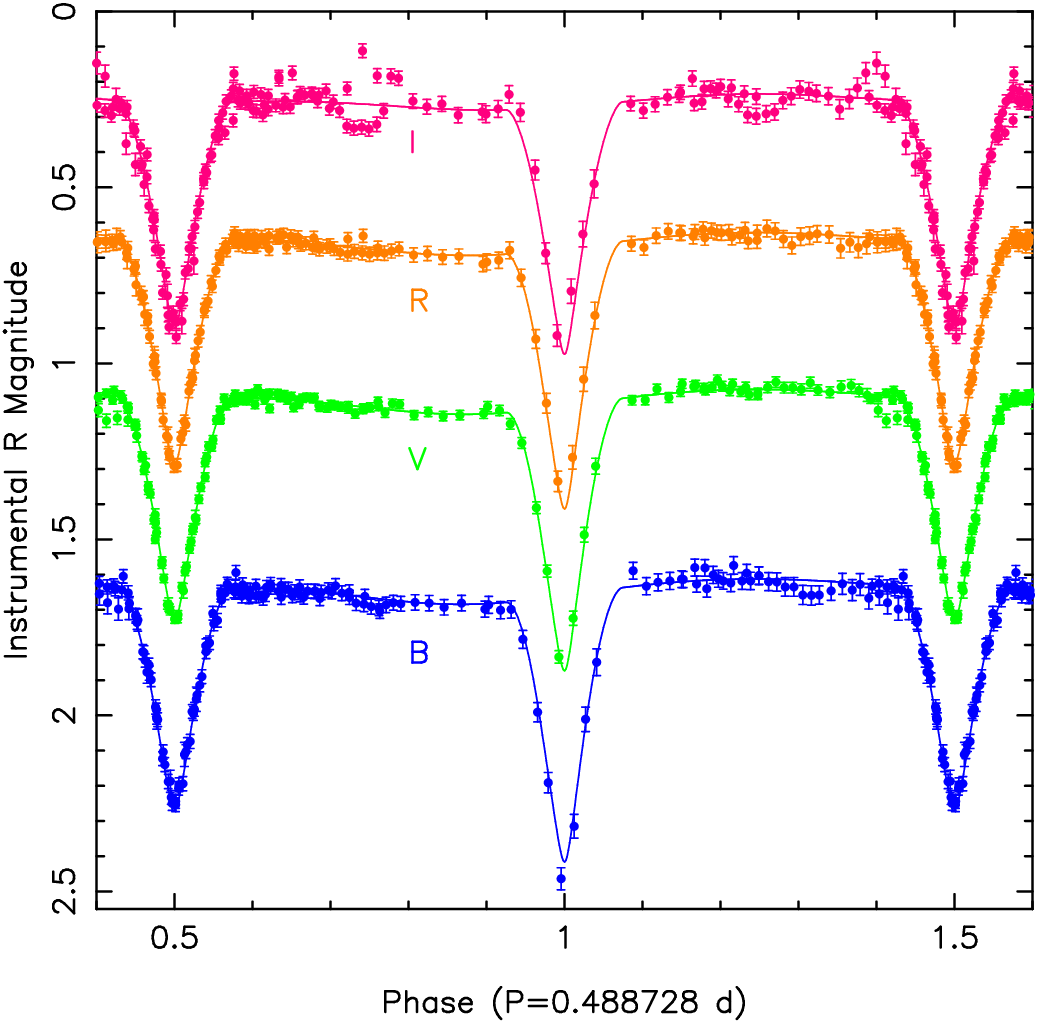}
\caption{The photoelectric light curves of GU Boo
obtained from Mount Laguna for the $V$,
$R$, and $I$ filters and the best-fitting ELC model.
The $I$-band light curve is somewhat noisy owing to a
much higher level of background light.\label{PMTlc}}
\end{figure}

\begin{figure}
\epsscale{.75}
\plotone{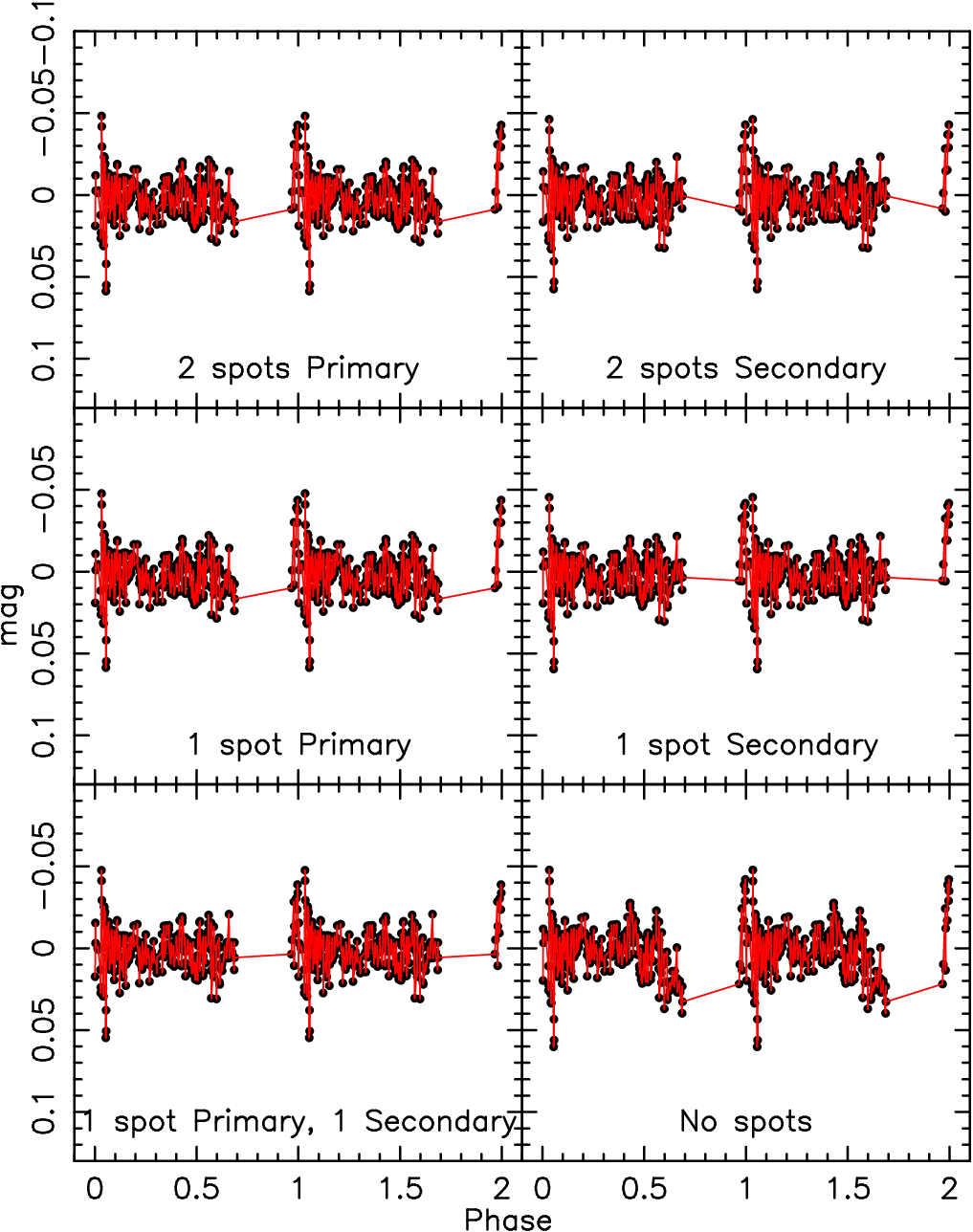}
\caption{The residuals of the model fits to the $R$-band CCD data for
the various spot scenarios:  Two spots on the primary star and
none on the secondary star (upper left),
two spots on the secondary star and none on the primary star (upper right),
one spot on the primary  and none on the secondary (middle left),
one spot on the secondary and none on the primary (middle right), one
spot on the primary and one spot on the secondary (lower left), and
no spots on either star (lower right).
\label{CCDres}}
\end{figure}

\begin{figure}
\epsscale{.75}
\plotone{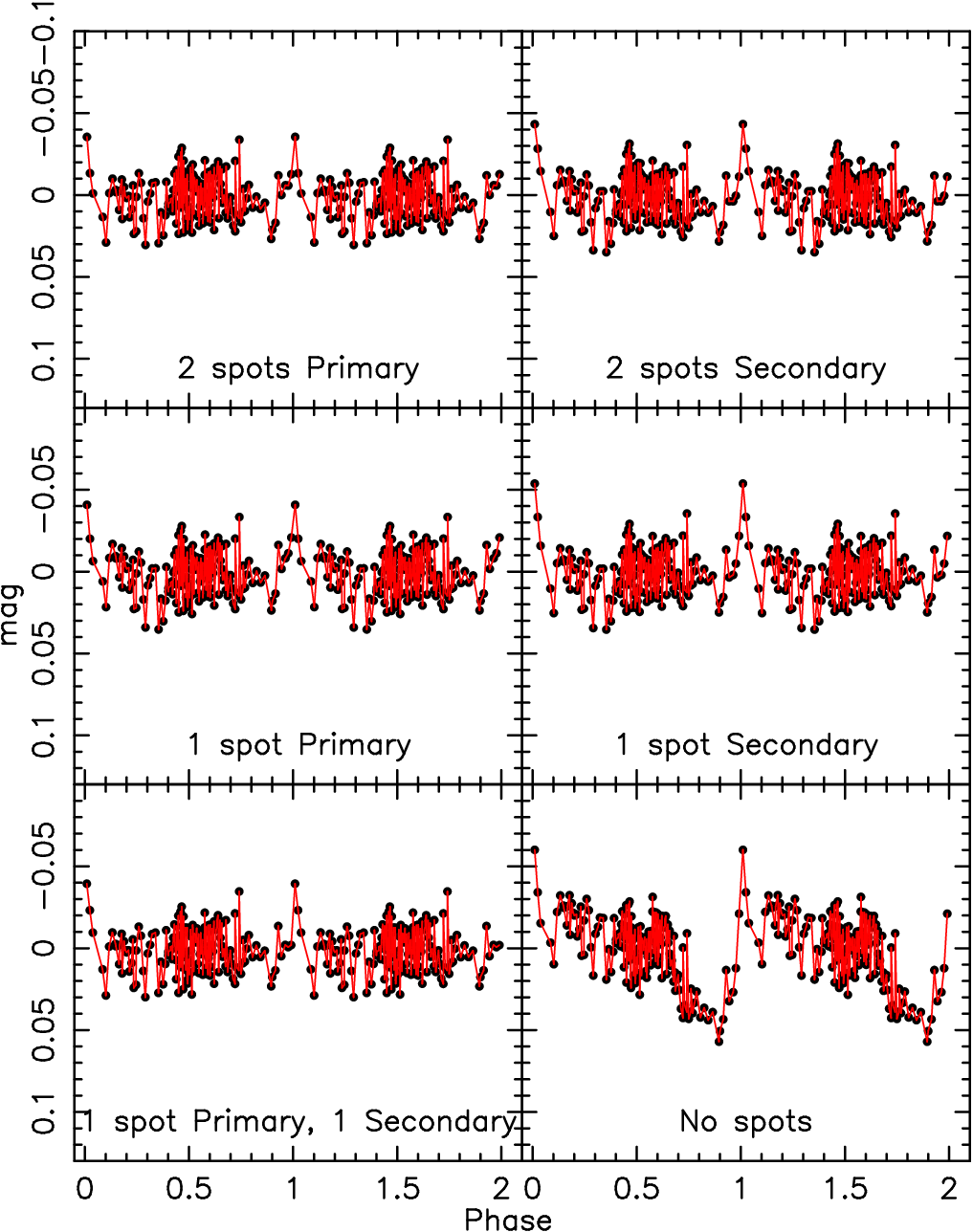}
\caption{Similar to Figure \protect{\ref{CCDres}}, but for the
$R$-band PMT light curve.
\label{PMTres}}
\end{figure}

\begin{figure}
\epsscale{.75}
\plotone{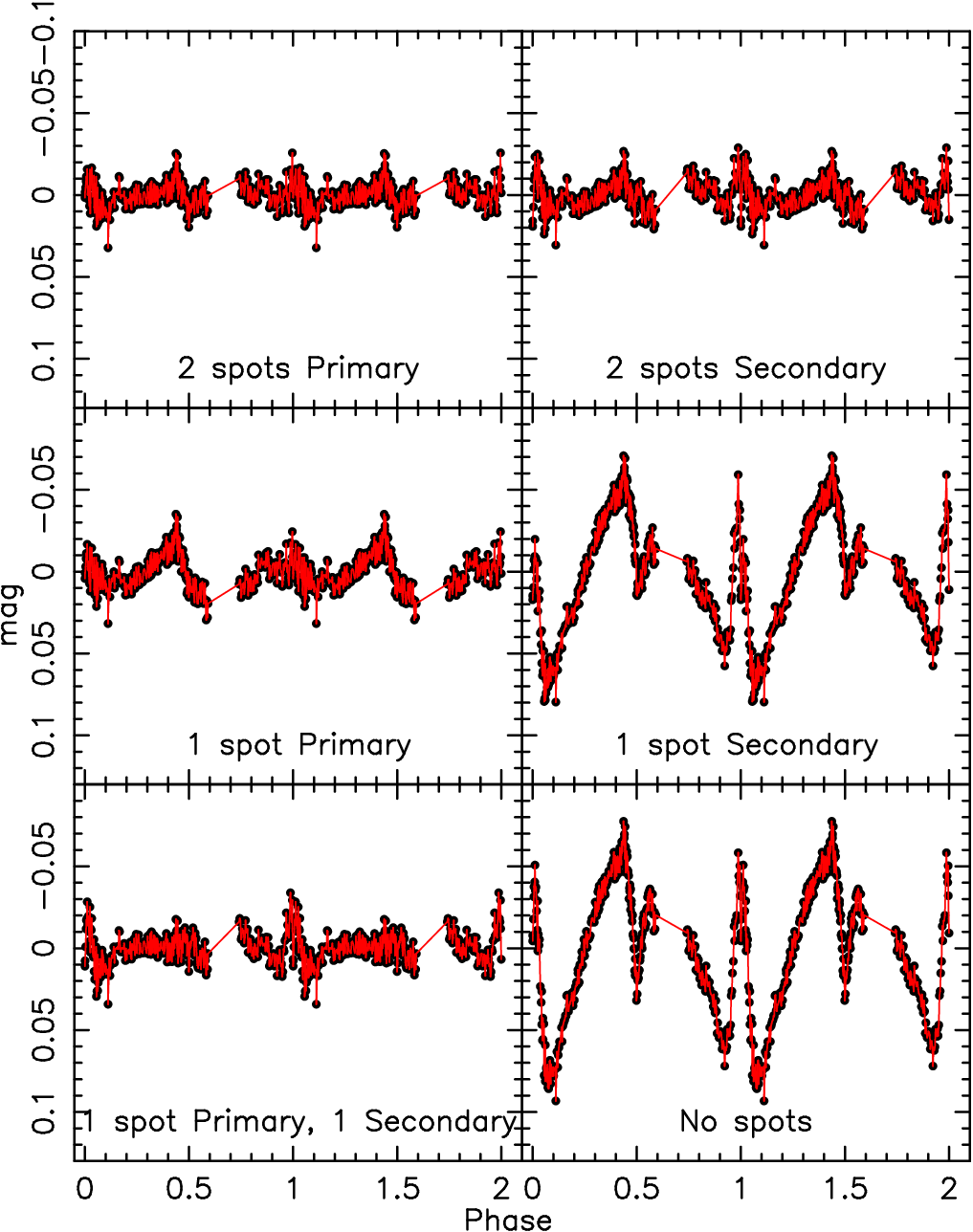}
\caption{Similar to Figure \protect{\ref{CCDres}}, but for the
$R$-band light curve from LR05.
\label{LRres}}
\end{figure}

The astrophysical parameters for GU Boo that are currently of most
interest to us are the masses and radii of the component stars.
In Table \ref{tab2}, we summarize the masses and radii derived from
the various data sets (CCD, PMT, and LR05) using the various spot
scenarios (no spots,
one spot on primary, one spot on secondary, one spot on
each, two spots on primary, and two spots on secondary).  For each
situation, we give the $\chi^2$ of the fit (which includes all
light curves in the particular data set and both radial velocity
curves), the component masses, the component radii, and the differences
between our derived values and the values reported in LR05.
One can use the $\chi^2$ values to determine which spot scenario provides
the optimal fit.  We find that for the CCD and PMT data,
the scenario with one spot on the primary and one spot
on the secondary gives the best fit. For the LR05 data,
two spots on the primary is optimal.
Table \ref{tab3} gives the input parameters for the best-fitting
models for each data set. Also given in Table \ref{tab3} are the derived
rotational velocities of each star. In all cases, the derived values agree
with the measured values given in LR05
($V_{\rm rot}\sin(i)=65$ and $58$ km s$^{-1}$ for the primary and
secondary, respectively, with
no uncertainty given).
Figures \ref{CCDlc} and \ref{PMTlc} show the phased CCD and PMT light
curves from Mount Laguna and the best-fitting models.
The agreement between the model
curves and the observed points is in general very good.  Figures
\ref{CCDres}, \ref{PMTres}, and \ref{LRres} show the $R$-band
residuals for the various spot scenarios for the CCD, PMT, and LR05
data, respectively.
For the Mount Laguna data,
it is hard to tell by eye the differences between the residuals from
the various spot scenarios (with the exception of the cases with no spots),
in spite of the fact that change in $\chi^2$ from the worst case to the best
case is significant.  On the other hand, the two-spot models are
clearly superior to the one-spot models for the LR05 data (Figure
\ref{LRres}).

We note some interesting features and trends seen in Table \ref{tab2}
and in Figures \ref{CCDres}, \ref{PMTres} and \ref{LRres}.  As one
might expect, the masses found the fits to the various data sets and
the various spot scenarios are very similar since all fits used the
same radial velocity curves.  A bit more surprising is the fact that
the radius of the primary and the radius of the secondary found from
the various data sets and spot scenarios generally agree with each
other at the $\approx 4\%$ level, although differences of up to about
8\% occur for a few of the cases.  With one or two exceptions, the
radii we found were within $\approx 0.02\,R_{\odot}$ of the values
reported in LR05\footnote{Since LR05 used the W-D code to model
their light curve, some of the differences seen in Table 2
might be due to differences in the modelling approach
and in the codes themselves}.
This is in spite of the fact that the change in
$\chi^2$ between the worst spot scenario and the best spot scenario
for a given data set is large, as noted above.  This would seem to
suggest that the radii one finds from the light curves is mostly
determined by the shapes of the eclipse profiles, and would be within
4 to 5\% of the ``true'' answer in most cases.  The spots can further
reduce the $\chi^2$ of the fit, but seem to add little in terms of the
radius determination.

On the other hand, mass and radius determinations at the 2\% level or
better are needed if one wants to perform detailed comparisons between
the measurements and the predicted values from evolutionary models.
Thus, for a given data set, one wants to have model light curves that
are well matched to the observed light curves.  As noted above, for
each data set, we found the spot scenario that resulted in the optimal
fit.  We summarize in Table \ref{tab4}
the masses and radii of the components
found from these models. We derived a radius of the primary of
$0.6329\pm 0.0026\,R_{\odot}$, $0.6413\pm 0.0049\,R_{\odot}$, and
$0.6373\pm 0029\,R_{\odot}$ from the CCD, PMT, and LR05 data,
respectively.  These values agree with the value reported by LR05
($R_1=0.623\pm0.016\,R_{\odot}$) at the level of $\approx 2\%$.  For
the secondary star, we derive radii $0.6074\pm 0.0035\,R_{\odot}$,
$0.5944\pm 0.0069\,R_{\odot}$, and $0.5976\pm 0.0059\,R_{\odot}$ from
the three respective data sets.  The LR05 value is $R_2=0.620\pm 0.020
\,R_{\odot}$.
In this case, our derived radii are all smaller than the LR05 values,
with the largest deviation being $\approx 3.5\%$.
Although the formal fitting errors are relatively
small ($\approx 0.5\%$ for the primary and $\approx 1\%$ for the
secondary), it seems that, for a given data set, the accuracy to which
we can determine the radii is limited to the
$\approx 2-4$\% level.  Since we
don't know the ``true'' radii of the stars in the GU Boo binary, it is
not immediately obvious if the presence of spots causes us to
overestimate slightly the radii or to underestimate slightly the
radii.  Therefore, taking an average of the three measurements may not
bring us closer to the true answer. Extensive Monte Carlo simulations
with model binaries might shed some light on this issue.

By now, it is well known that evolutionary models for low mass stars have
done a relatively poor job when confronted with mass-radius measurements
from eclipsing binaries.  In spite of the fact that there may be systematic
errors of a few percent on the radius determinations,
the measured masses and radii of low mass stars in eclipsing binaries
are significantly larger than those predicted based
on evolutionary models (e.g., Torres \& Ribas 2002;
LR05, Bayless \& Orosz 2006, Ribas 2006).
Recently, there has been speculation that
unusually strong stellar activity in
these low-mass stars
might make them larger than they otherwise would be
(Ribas 2006; Torres et al.\ 2006;
L\'opez-Morales 2007; Chabrier et al.\ 2007), and that the changes in the
stars caused by stellar activity has not been properly accounted
for in the evolutionary models.
As discussed here, one or both of the stars in GU Boo have large spots
that change with time, and these spots seem to limit our ability to
derive radii accurate at the level of $\lesssim 2\%$.
Nearly all
of the well-studied eclipsing binaries with low-mass stars
have orbital periods shorter than $\approx 3$ days.  Since the timescale
for tidal synchronoziation for these binaries is relatively short
(e.g., Zahn 1977), the stars presumably have short rotation periods
and higher amounts of activity compared to single stars of similar
mass.
A recent exception is a binary known as
T-Lyr1-17236 (Devor et al.\ 2008), which has an orbital period of
about 8.43 days and components with masses and radii of
$M_1=0.6795\pm 0.0107\,M_{\odot}$,
$M_2=0.5226\pm 0.0061\,M_{\odot}$,
$R_1=0.634\pm 0.043\,R_{\odot}$, and
$R_2=0.525\pm 0.052\,R_{\odot}$, respectively.  Both stars have relatively
small rotational velocities, and
with such a long orbital
period, would still be slowly rotating even if the binary has been
circularized because of tidal forces. Devor et al.\ (2008) show there
are no obviously strong indicators of stellar activity in these stars.
Although the radius measurements have relatively large uncertainties,
Devor et al.\ (2008) show that both stars have radii consistent with
predictions based on evolutionary models.

If the stellar activity is indeed the cause of the disagreement
between the measured radii and the radii predicted from evolutionary
models, then presumably there is a threshold below which
the activity has little or no effect on the overall structure of the
star.  We have shown (for GU Boo at least) that
star spots seem to be a
limiting factor in an accurate radius determination,
and likewise one would expect that there is also a threshold of
spot activity below which the radius determination can become
much more precise.  A better observational
understanding of the former threshold can come from the study of
additional long-period binaries.  Although these binaries are rare,
hopefully more will be discovered in current and future large area surveys
(for example the Trans-Atlantic Exoplanet Survey [TrES,
Alonso et al.\ 2004] that led to
the discovery of T-Lyr-17236).
A better observational understanding of
the latter can come from long-term monitoring of the known systems.
As we have done for GU Boo, one can observe these binaries at different
times and derive radii from independent light curves.  The different
measurements will have a spread, either large or small, and the size
of the spread might be correlated with the level of spot activity.

\acknowledgments

We thank Philip Rosenfield, Leah Huk, David Garcia,
and Chad Downum for their assistance with the
observations at Mount Laguna Observatory.

\clearpage

\clearpage

\begin{deluxetable}{lrrrrr}
\tabletypesize{\scriptsize}
\tablecaption{Observations by Dataset, Date, and Band\label{tab1}}
\tablewidth{0pt}
\tablehead{
\tableline
\colhead{Data Type} & \colhead{UT date} &
\colhead{$N_{B}$}\tablenotemark{a} & \colhead{$N_{V}$} & \colhead{$N_{R}$} & \colhead{$N_{I}$}
}

\startdata
\rm LR05, Kitt Peak        & 2003-Mar-25 & \nodata & \nodata & \nodata &     190 \\
                           & 2003-Apr-02 & \nodata & \nodata & \nodata &     178 \\
                           & 2003-Apr-03 & \nodata & \nodata &     107 & \nodata \\
                           & 2003-Apr-19 & \nodata & \nodata &     163 & \nodata \\
                           & 2003-May-02 & \nodata & \nodata &      95 &     254 \\
\tableline
\rm CCD, Mount Laguna      & 2005-May-26 & \nodata &      35 &      36 &      34 \\
                           & 2005-May-27 & \nodata &     125 &     121 &     121 \\
                           & 2005-May-28 & \nodata &     135 &     139 &     131 \\
\tableline
\rm PMT, Mount Laguna      & 2005-Jun-02 &      13 &      13 &      13 &       13 \\
                           & 2005-Jun-03 &       2 &       2 &       2 &        2 \\
                           & 2005-Jun-04 &      42 &      42 &      42 &       42 \\
                           & 2005-Jun-05 &      23 &      23 &      23 &       23 \\
                           & 2005-Jun-06 &      39 &      39 &      39 &       39 \\
                           & 2005-Jun-18 &      30 &      30 &      30 &       30 \\
                           & 2005-Jun-19 &      11 &      11 &      11 &       11 \\
\tablenotetext{a}{Number of data points in the \rm{B,V,R, and I} bands respectively }
\enddata

\end{deluxetable}

\begin{deluxetable}{clrrrrrrrrr}
\tabletypesize{\scriptsize}
\rotate
\tablecaption{One and Two Spots Optimization Results\label{tab2}}
\tablewidth{0pt}
\tablehead{
\colhead{Data~Type} & \colhead{Spots} & \colhead{$\chi^{2}$} &
\colhead{$M_{1}$\tablenotemark{a}} & \colhead{$M_{2}$\tablenotemark{b}} & \colhead{$R_{1}$\tablenotemark{c}} &
\colhead{$R_{2}$} & \colhead{$\Delta M_{1}$\tablenotemark{d}} &
\colhead{$\Delta M_{2}$} & \colhead{$\Delta R_{1}$\tablenotemark{e}} &
\colhead{$\Delta R_{2}$} }
\startdata

CCD & 1~Primary & 897.602 & 0.6111 & 0.6006 & 0.6316 & 0.6224 & 0.001106 &
0.001581 & 0.008593 & 0.002363 \\

CCD & 1~Secondary & 893.771 & 0.6188 & 0.6058 & 0.6359 & 0.6333 & 0.0088 &
0.0068 & 0.0129 & 0.0133 \\

CCD & 2~Primary & 890.460 & 0.6115 & 0.6006 & 0.6296 & 0.6214 & 0.001505 &
0.001595 & 0.006576 & 0.001406 \\

CCD & 2~Secondary & 872.868 & 0.6191 & 0.6054 & 0.6473 & 0.6331 & 0.0091 &
0.0064 & 0.0243 & 0.0131 \\

CCD & 1~Prim~~1~Sec & 854.273 & 0.6049 & 0.5932 & 0.6329 & 0.6074 & $-0.0051$ &
$-0.0058$ & 0.0099 & $-0.0126$ \\

CCD & No Spots\tablenotemark{f} & 995.440 & 0.6124 & 0.6027 & 0.6273 & 0.6579 &
0.0024 & 0.0037 & 0.0043 & 0.0379 \\

PMT & 1~Primary & 1080.768 & 0.6107 & 0.5963 & 0.6429 & 0.5979 & 0.000724 &
$-0.002664$ & 0.019905 & $-0.022113$ \\

PMT & 1~Secondary & 1088.551 & 0.6245 & 0.6105 & 0.6459 & 0.6038 & 0.0145 &
0.0115 & 0.0229 & $-0.0162$ \\

PMT & 2~Primary & 1055.538 & 0.6073 & 0.5936 & 0.6413 & 0.5944 & $-0.0027$ &
$-0.0054$ & 0.0183 & $-0.0256$ \\

PMT & 2~Secondary & 1053.480 & 0.6188 & 0.6059 & 0.6406 & 0.6121 & 0.008756 &
0.006937 & 0.017633 & $-0.007892$ \\

PMT & 1~Prim~~1~Sec & 1046.102 & 0.6013 & 0.5887 & 0.6195 & 0.6036 & $-0.0087$ &
$-0.0103$ & $-0.0035$ & $-0.0164$ \\

PMT & No~Spots \tablenotemark{f} & 1548.052 & 0.6118 & 0.6026 & 0.6226 &
0.5928 & 0.0018 & 0.0036 & $-0.0004$ & $-0.0272$ \\

\tableline
Other Data & & & & & & & & & & \\
\tableline

LR05\tablenotemark{g} & 1~Primary & 1346.136 & 0.5983 & 0.5797 & 0.6286 & 0.6107 &
$-0.0117$ & $-0.0193$ & 0.0056 & $-0.0093$ \\

LR05 & 1~Secondary & 1657.060 & 0.6223 & 0.6033 & 0.6644 & 0.6430 &     0.0123
& 0.0043 & 0.0414 & 0.0230 \\

LR05 & 2~Primary & 1052.124 & 0.6002 & 0.5847 & 0.6373 & 0.5976 & $-0.0098$ &
$-0.0143$ & 0.0143 & $-0.0224$ \\

LR05 & 2~Secondary & 1134.496 & 0.6301 & 0.6126 & 0.6292 & 0.6284 & 0.0201 &
0.0136 & 0.0062 & 0.0084 \\

LR05 & 1~Prim~~1~Sec & 1115.721 & 0.6268 &      0.6053 & 0.6448 & 0.6031 &
0.0168 & 0.0063 & 0.0218 & $-0.0169$ \\

LR05 & No~Spots \tablenotemark{f} & 8814.927 & 0.6098 & 0.6068 & 0.6055 &
0.6728 & $-0.0002$ & 0.0078 & $-0.0175$ & 0.0528 \\

\enddata
\tablenotetext{a}{Mass of Primary in $M_{\odot}$ units}
\tablenotetext{b}{Mass of Secondary in $M_{\odot}$ units}
\tablenotetext{c}{Radius of Primary in $\rm{R_{\odot}}$ units}
\tablenotetext{d}{$\Delta M_{1}=M_{1~\rm fit} - M_{1~\rm published~in~LR05}$}
\tablenotetext{e}{$\Delta R_{1}=R_{1~\rm fit} - R_{1~\rm published~in~LR05}$}
\tablenotetext{f}{For the given data set, this is the best optimized solution
that includes no spots. When the spots are simply removed from the best solution in
the given case, the resultant $\chi^{2}$ values are 1367.048, 1639.882,
12541.381 for the CCD,PMT, LR05~ data sets, respectively}
\tablenotetext{g}{Optimization using data from  LR05}
\end{deluxetable}

\begin{deluxetable}{crrrrrrr}
\tabletypesize{\scriptsize}
\tablecaption{All Fitted Parameters of the Best Solutions\label{tab3}}
\tablewidth{0pt}
\tablehead{
Data~Type: & \multicolumn{2}{c}{\rm CCD, Spots: 1~Prim' 1~Sec'} &
\multicolumn{2}{c}{\rm PMT, Spots: 1~Prim' 1~Sec'} & \multicolumn{2}{c}{\rm LR05, Spots: 2~Primary}\\
\tableline
\colhead{Parameter} & \colhead{Value} & \colhead{Uncertainty} &
\colhead{Value} & \colhead{Uncertainty} & \colhead{Value} &
\colhead{Uncertainty} & \colhead{Unit}
}

\startdata
$\chi^{2}$             & 854.2734 & \nodata &  1046.1019 & \nodata & 1052.1241 & \nodata & \\
$M_{1}$                & 0.6049 & $\pm 0.00489$ \tablenotemark{a} & 0.6014  & $\pm 0.0106$ & 0.6002 & $\pm 0.0060$ & \rm {$M_{\odot}$} \\
$R_{1}$                & 0.6329 & $\pm 0.00261$ & 0.6195  & $\pm 0.0077$ & 0.6373  & $\pm 0.0029$  & \rm {$R_{\odot}$} \\
$\frac{R_{1}}{R_{2}}$  & 1.0419 &  $\pm 0.007663$ & 1.0264 & $ \pm 0.0161$ & 1.0666 & $\pm 0.0153$ \\
$i$                    & 88.2804 &  $\pm 0.1433$ & 88.0500 & $\pm 0.2533$ & 88.6340 & $\pm 0.1749$ & \rm {deg} \\
$K_{1}$                & 142.0709 &  $\pm 0.7019$ & 141.6102  & $\pm 0.7922$ & 141.1003 & $\pm 1.1165$ & $\rm{km~sec^{-1}}$ \\
$T_{1}$                & 3737.7100 &  $\pm 12.24$ & 3701.1500 & $\pm 29.4800$ & 3788.5100 & $\pm 6.99$ & $ \rm {^{\circ}K}$ \\
$T_{2}$                & 3625.8300 &  $\pm 14.37$ & 3625.5500 & $\pm 31.2600$ & 3706.3400 & $\pm 9.9$ & $ \rm {^{\circ}K}$ \\
$P$                    & 0.48873066 &  $\pm 1.5 \times 10^{-7}$ & 0.488730245 & $\pm 2.9 \times 10^{-7}$ & 0.488718 & $\pm 2.165 \times 10^{-5}$ & \rm {day} \\
$T_{0}$                & 2723.9811 &  $\pm 0.0002108$ & 2723.9816 & $\pm 0.000492$ & 2723.9856 & $\pm 0.00535$ & \rm {HJD} \\
$b1$ \tablenotemark{b} & 0.8052 &  $\pm 0.0286$ & 0.9476 & $\pm 0.0196$ & 0.9256 & $\pm 0.0062$ & \\ 
$b2$ \tablenotemark{c} & 10.7800 &  $\pm 0.4$ & 33.7500 & $\pm 4.23$ & 46.4600 & $\pm 2.4900$ & \rm {deg} \\
$b3$ \tablenotemark{d} & 59.0700 &  $\pm 1.38$ & 57.8000 & $\pm 2.6900$ & 353.1000 & $\pm 1.0100$ &\rm {deg} \\
$b4$ \tablenotemark{e} & 56.7140 &  $\pm 0.522$ & 57.9220 & $\pm 5.5700$ & 38.2680 & $\pm 2.0810$  & \rm {deg} \\
$c1$ \tablenotemark{f} & 0.7539 &  $\pm 0.0567$ & 0.5543  & $\pm 0.1534$ & 1.1673 & $\pm 0.0192$ & \\ 
$c2$                   & 42.5800 &  $\pm 2.91$ & 22.7700 & $\pm 3.0100$ & 104.9300 & $\pm 26.9000$ & \rm {deg} \\
$c3$                   & 54.4400 &  $\pm 3.16$ & 48.2200 & $\pm 3.6800$ &207.5500 & $\pm 2.8600$ & \rm {deg} \\
$c4$                   & 18.0550 &  $\pm 1.806$ & 28.8490 & $\pm 4.2680$ & 9.8230 & $\pm 0.7860$ & \rm {deg} \\
\tableline
$V\sin(i)_{1}$\tablenotemark{g} & 65.5090  &  $\pm 0.2670$ & 64.1180  & $\pm 0.7910$ & 65.9850 & $\pm 0.3050$ & $\rm{km~sec^{-1}}$ \\
$V\sin(i)_{2}$                  & 65.8790  &  $\pm 0.3570$ & 62.4730  & $\pm 1.3100$ & 61.8700 & $\pm 0.606$ & $\rm{km~sec^{-1}}$ \\

\enddata
\tablenotetext{a}{All uncertainties are calculated using
parameter values at $\chi^{2} \lesssim \chi^{2}_{\rm min}+2$}
\tablenotetext{b}{spot Temperature factor}
\tablenotetext{c}{spot Latitude}
\tablenotetext{d}{spot Longitude}
\tablenotetext{e}{spot Angular Radius}
\tablenotetext{f}{c1-c4 are similar to b1-b4 but are for the spot on the
 secondary; when both spots are on the same star, such as in the LR05 case, the
 parameters for the second spot are tabulated b5-b8}
\tablenotetext{g}{derived rotational velocities for the primary and secondary}

\end{deluxetable}

\begin{deluxetable}{lrrrr}
\tabletypesize{\scriptsize}
\tablecaption{Masses and Radii from All Best Solutions\label{tab4}}
\tablewidth{0pt}
\tablehead{
\tableline
\colhead{Data Type} & \colhead{$R_{1}$\tablenotemark{a}} &
\colhead{$R_{2}$ \tablenotemark{b}} & \colhead{$M_{1}$ \tablenotemark{c}} &
\colhead{$M_{2}$}
}

\startdata
\rm CCD (spots: 1~Primary~1~Secondary) & $0.6329 \pm 0.0026$ & $0.6074 \pm 0.0035$ & $0.6049 \pm 0.0049$ & $0.5932 \pm 0.0062$ \\
\rm PMT (spots: 1~Primary~1~Secondary) & $0.6413 \pm 0.0049$ & $0.5944 \pm 0.0068$ & $0.6073 \pm 0.0063$ & $0.5936 \pm 0.0033$ \\
\rm LR05 (spots: 2~Primary)            & $0.6373 \pm 0.0029$ & $0.5976 \pm 0.0059$ & $0.6002 \pm 0.0060$ & $0.5847 \pm 0.0090$\\
\multicolumn{1}{c}{Average \tablenotemark{d}} &
\multicolumn{1}{r}{0.6372 $\pm 0.0042$} & \multicolumn{1}{l}{0.5998 $\pm 0.0068$} &
\multicolumn{1}{l}{0.6041 $\pm 0.0036$} & \multicolumn{1}{l}{0.5905 $\pm 0.0050$}\\
\rm LR05 own fit (spots: 1~Primary~1~Secondary) & $0.6230 \pm 0.0160$ & $0.6200 \pm 0.0200$ & $0.6100 \pm 0.0070$ & $0.5990 \pm 0.0060$\\

\tablenotetext{a}{Radius of the primary in $R_{\odot}$}
\tablenotetext{b}{Radius of the secondary in $R_{\odot}$}
\tablenotetext{c}{Mass of the primary in $M_{\odot}$}
\tablenotetext{d}{The uncertainty of the average is taken to be the
                 standard deviation in the values of the given parameter}

\enddata

\end{deluxetable}

\end{document}